\documentclass[12pt,a4paper,oneside]{article}
\usepackage{graphicx}
\begin{document}

\begin{titlepage}
\hfill{\small PRA-HEP 99/04}
\\
\vskip2cm
\begin{center}
\Large\bf Operator product expansion and analyticity
\end{center}

\vspace{1.0cm}

\begin{center}
Jan Fischer\footnote{e-mails: fischer@fzu.cz, Jan.Fischer@cern.ch}\\
Institute of Physics, Academy of Sciences of the Czech Republic,\\ 
CZ-182 21  Prague 8, Czech Republic
\end{center}
\centerline{and}

\begin{center}
Ivo Vrko\v{c}\footnote{e-mail: vrkoc@matsrv.math.cas.cz}\\ 
Mathematical Institute, Academy of Sciences of the Czech Republic, \\
\v{Z}itn\'{a} 25, CZ-115 67  Prague 1,  Czech Republic
\end{center}
\vspace{1.2cm}
\begin{abstract}
We discuss the current use of the operator-product expansion in QCD 
calculations. Treating the OPE as an expansion in inverse powers of 
an energy-squared variable (with possible exponential terms added), 
approximating the vacuum expectation value of the operator product by 
several terms and assuming a bound on the remainder along the euclidean 
region, we observe how the bound varies with increasing deflection from 
the euclidean ray down to the cut (Minkowski region). We argue that the 
assumption that the remainder is constant for all angles in the cut complex 
plane down to the Minkowski region is not justified. 

Making specific assumptions on the properties of the expanded function, we 
obtain bounds on the remainder in explicit form and show that they are very 
sensitive both to the deflection angle and to the class of functions 
considered. The results obtained are discussed in connection with 
calculations of the coupling constant $\alpha_{s}$ from 
the $\tau$ decay.  
\vspace{0.6cm}

 
\vspace*{0.6cm}

\noindent July 1999\hfill
\end{abstract}
\end{titlepage}
\newpage
\section{Introduction}

The operator-product expansion (OPE) \cite{Wilson, Zim}, 
\begin{equation}
{\rm i} \int {\rm d}x {\rm e}^{{\rm i}qx}A(x)B(0) \approx
\sum_{k} C_{k}(q){\cal O}_{k} ,
\label{OPEq}
\end{equation} 
represents the product $AB$ of two local operators as a combination, 
with {\it c}-number coefficients, of the local operators 
${\cal O}_{k}$. Here, $q$ is the total four-momentum of the system 
considered and $q^{2}=s=-Q^2$. The singularities of the product 
are contained in the coefficient functions $C_{k}(q)$, which 
are ordered according to the increasing exponent $k$ in $s^{-k}$.

In local quantum field theory, the product $A_{1}(x^{1})...A_{a}(x^{a})$ 
of two or several field operators is singular for coinciding arguments, 
and the problem of defining it in a neighbourhood is of fundamental 
importance. K. Wilson proposed that the operator product may be expanded 
in the form
\begin{equation}
A_{1}(x+x^{1})...A_{a}(x+x^{a}) = \sum_{k=0}^{n}f_{k}(x^{1},...,x^{a})
{\cal O}_{k}(x) + R_{n}(x,x^{1},...,x^{a}),
\label{Zim1}
\end{equation}
where the remainder $R_{n}$ vanishes in the limit $x^{j} \rightarrow 0$ 
while the functions $f_{k}$ become singular or non-vanishing. He also
generalized his hypothesis  by assuming \cite{Wilson} that any operator 
product may be represented as a series
\begin{equation}
A_{1}(x+x^{1})...A_{a}(x+x^{a}) = 
\sum_{k=0}^{\infty}f_{k}(x^{1},...,x^{a}){\cal O}_{k}(x) , 
\label{Zim2}
\end{equation}
which is asymptotic in the sense that to every $N$ there 
exists a $k(N)$ such that the coefficients $f_{k}(\rho \lambda^{1},..., 
\rho \lambda^{a})$ vanish faster than $\rho^{N}$ for all $k\geq k(N)$.  

A rigorous formulation of the operator-product expansion within perturbation
theory has been worked out by Zimmermann \cite{Zim}. The Wilson relation can be
justified order by order in perturbation theory, and there is a well-defined
algorithm allowing one to calculate the coefficient functions. Zimmermann, 
Wilson and Otterson \cite{WZ} show how an operator product expansion can be 
derived from general principles, and find conditions under which the OPE gives 
complete information on the short-distance behaviour of operator products. 

Ferrara et al. \cite{Fer} examine how the form of the operator-product
expansion depends on the symmetry group of the theory. They find for 
instance that the covariance under the spinor group SU(2,2) can place 
significant restrictions on the structure of the expansion terms on the 
light cone. 

In momentum space, the Minkowski region lies along the cut $Q^{2}<0$ 
carrying the spectrum of physical states, while in the euclidean 
region ($Q^{2}>0$) the expansion is, for large $Q^2$, determined 
by short-distance dynamics, the separation of the large-distance 
contributions from the short-distance ones being well defined \cite{Shif}. 
Predictions in the Minkowski region are obtained by analytic continuation. 

The operator product expansion has been applied to various problems in 
quantum theory with varying degree of rigour. According to the problem 
considered, different mathematical properties of the expansion have been  
proved or assumed (also the symbol $\approx$ in (\ref{OPEq}) is understood 
differently in different contexts). At a fixed perturbative order, one can 
express the operator product in terms of the Feynman diagrams of this order. 
There are reasons to believe that the large-momentum expansion has a 
non-vanishing convergence radius. Smirnov \cite{Smir1} (see also 
\cite{Smir2} and references therein) proved asymptotic expansions 
of the renormalized Feynman amplitudes in the large-momentum 
(mass) limit, and found the corresponding operator 
expansions for the S-matrix and composite operators. 

In quantum chromodynamics, a theory with a strong non-perturbative component, 
very little is known about the mathematical character of the operator-product 
expansion and its exact composition. In particular, it is not known whether 
terms exponential in the variable $(Q^2)^{1/2}/\Lambda$, for instance terms 
of the form
\cite{Shif} 
\begin{equation}
\exp (-C(Q^2)^{1/2}/\Lambda)
\label{Shif}
\end{equation} 
with $C$ positive, have to be added to describe strong-interaction processes. 
There are reasons to believe that the series is divergent, but it is not 
known whether it is asymptotic to the function searched for and, if so, in 
what region of the $Q^2$ plane. We can expect that there is more chance to 
trust (\ref{OPEq}) at higher energies than at low energies; but the 
large-order behaviour of its terms is not known \cite{Ben}.  

In contrast to the lack of rigour, the operator-product expansion in QCD 
is very much needed as a tool for solving a number of practical problems, 
such as the semileptonic B-meson decay, heavy-light quark systems, heavy 
quarkonia and the Drell-Yan process. The inclusive decay hadronic widths 
are also expected to be calculable using the OPE and analyticity. 

To apply the operator product expansion in QCD, one is of course faced with 
the problem of extending it to non-perturbative dynamics. This issue 
has been discussed since the development of the QCD sum rules 
\cite{SVZ}, where OPE is combined with analyticity and other 
non-perturbative aspects of QCD. The successful application 
of the QCD sum rule technique to many processes 
and effects is well known.  

Secondly,  some information about the behaviour of the operator product 
expansion in the complex $Q^2$ plane away from euclidean region, along all 
rays passing through the origin, is necessary. The reason is that the 
Feynman graphs, through which the observables (including those describing 
high-energy effects) are expressed, contain integration over small 
momenta, where there is little chance that the OPE can be applied. 
Using the analyticity property, however, one can replace the 
low-energy integral by that along a circle of a sufficiently 
large radius in the complex plane (see Fig. 1).

\begin{figure}[htb]
  \begin{center}
    \includegraphics*[width=8cm]{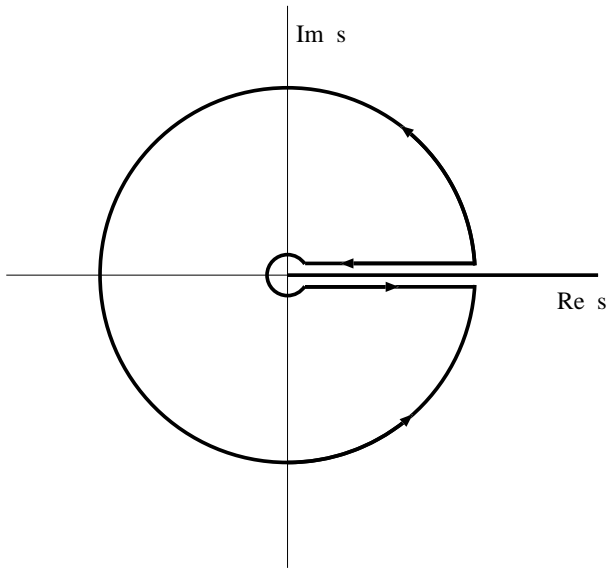}
    \caption{Integration contour in the $s=-Q^2$ plane, showing how analyticity 
    is used to replace an integral along the cut by that along a circle. 
     For the decay of the $\tau$ lepton (see section 3), the radius is equal 
     to its mass $m_{\tau}$ squared.} 
    \label{fig:t}
  \end{center}
\end{figure}

 This, however, poses a new problem, that of finding conditions under 
 which an expansion of the type 
\begin{equation}
f(1/Q^2) \approx \sum_{k}a_{k}(q)/Q^{2k} 
\label{approx}
\end{equation}
can be extended to the complex $Q^2$ plane, to be valid along all rays 
$|Q^2| \rightarrow \infty$. Such an extension is a delicate problem 
requiring precise mathematical conditions, which are not known in 
the case of QCD. To make the problem well-defined, recourse to 
simplifying mathematical assumptions therefore seems necessary. 

While Wilson's operator product expansion is originally formulated in the
Euclidean domain, its applications are mostly related to quantities of the
Minkowski nature. Then, the assumption is usually adopted \cite{BNP} 
that the ``convergence properties'' of the operator-product expansion 
away from the euclidean ray $Q^2>0$ are the same as those along it, except 
the points of the cut $Q^2<0$ (the Minkowski region). Simultaneously, it is 
assumed that the truncation error (caused by approximating the operator 
product by several terms of the expansion) is {\it independent} of the 
direction in the complex $Q^2$ plane (see Fig. 1). 

We consider this assumption too a severe simplification, in addition 
technically motivated: indeed, it is hard to believe that the unknown 
discontinuity along the cut would have no influence on the behaviour 
of the function along rays that are near the cut. 

In the present paper we therefore look for a model scheme that would not a
priori exclude the possibility that the bound on the truncation error (assumed 
originally in euclidean region) becomes looser with increasing deflection from 
the positive real semiaxis. The program of the paper was sketched in our 
previous papers, see \cite{JaFi} and \cite{Mont}: We discuss conditions under 
which an expansion of the type (\ref{approx}) can be extended to angles away 
from the euclidean semiaxis in the complex $Q^2$ plane, with the aim to find 
how a bound, originally assumed to be valid along the positive real semiaxis, 
develops when the deflection increases and the cut is approached. We propose 
(in section 2 of the present paper) a scheme that ({\it i}) has precise 
mathematical meaning, ({\it ii}) is free from the {\it a priori} assumption 
that the influence of the cut on the truncation error along a general ray can 
be neglected and, simultaneously, ({\it iii}) tries to keep the model possibly 
close to real situations. To illustrate the physical interest of such a 
problem, we discuss in section 3 the determination of the coupling constant 
from the $\tau$ lepton hadronic width. Our results are presented and discussed 
in section 4. In the concluding section 5 we summarize our results, and 
discuss several possibilities of refining the scheme to get closer to 
physical reality. The three Appendices contain the proof of the main 
theorem on which our results are based, and an example of a function 
that saturates the bound (\ref{VrA3}) obtained in the Appendix A.

\section{The operator-product expansion away from\\
euclidean region}
Let $F(s)$ be holomorphic in ${\cal C}$, the complex $s$ plane 
($s=-Q^2$) cut along $s\,\in \,[0, \infty]$ from which a bounded
domain around the origin may be removed. Let the constants $a_{k}$, 
$k=0,1,2,...n-1$, and a positive number $A_{n}$ exist such that the 
following inequality
\begin{equation}
|F(s) - \sum_{k=0}^{n-1} a_{k}/(-s)^{k}| < A_{n}/|s|^n  
\label{V2}
\end{equation}
is satisfied for a positive integer $n$ and all $s<-b$, with $b$ being a 
positive number. (To avoid unnecessary complications, we take the lowest 
value of $k$ to be zero, assuming that terms that are infinite at
$|s|=\infty$ have been removed.) The problem is what inequality 
(if any) will hold  along rays in the complex $s$ plane, away 
from the negative real semiaxis (euclidean region). 

As is natural to expect, the answer depends on additional 
assumptions imposed on the function $F(s) \equiv f(1/Q^{2})$. 
When compared with the problem of the operator-product expansion in QCD, 
we make the following extra assumptions:     

1. The coefficient functions $a_{k}$ of local operators are perturbative
series in powers of the QCD coupling constant $\alpha_{s}(Q^2)$, which in turn
is expanded in negative powers of $\ln(Q/\Lambda)$ with coefficient functions
depending on $\ln \ln(Q/\Lambda), \ln \ln \ln(Q/\Lambda)$, etc. Here, $\Lambda$ 
is $\Lambda_{\rm QCD}$, the fundamental scale of quantum chromodynamics. 
It is sometimes instructive to consider the form of the power corrections in
the case that the logarithmic $Q^2$-dependence of the coefficient functions 
is neglected. We make this approximation for simplicity in the present paper, 
with the aim to examine the general case in a later publication. (In this 
approximation, the cut due to the logarithmic dependence of the Wilson 
coefficients disappears, but $F(s)$ can still have a cut if the 
expansion has an infinite number of terms or if additional singular 
terms are added to the series, see below.) 

2. We assume that the inequality (\ref{V2}) is valid up to $s = - \infty$ , 
where the right-hand side of (\ref{V2}) vanishes at a high rate. This 
amounts to assuming that the $n$-th order remainder $R_{n}(-1/s)$,  
\begin{equation} 
R_{n}(z) = f(z)-\sum_{k=0}^{n-1}a_{k} z^{k} , 
\label{rem}
\end{equation}
tends to zero for $z \rightarrow 0$ as the $n$-th power of $z$ for 
at least one value of $n$.\footnote{We use the notation  
$z = r{\rm e}^{{\rm i}\varphi}= x+{\rm i}y, z=-1/s$ and $f(z) \equiv F(s)$.} 

These two assumptions simplify our problem but they may move us farther  
from physics. We plan to refine the scheme in a subsequent paper; our 
principal intention here is to create a model which would be free from the 
conventional assumption that the truncation error is the same in all 
directions (except the cut). 
 
In phenomenological applications, the starting estimate of the truncation 
error in euclidean region is usually taken to be of the order of the first 
neglected term of the expansion. This term serves as a pragmatic guidance 
in estimating the size of the error; then, its value is conventionally 
extended to be the estimate of the truncation error along all rays 
passing through the origin. 

It is our aim to obtain a more realistic picture about how the error may 
develop when the cut is approached. We examine the high-energy properties 
of the expansion (\ref{approx}) along different rays in the complex plane. 
Assuming the bound (\ref{V2}) for $s<-b$, we observe how it varies with 
increasing deflection of the ray from euclidean region. 	

Note that we do not demand that the expansion (\ref{approx}) be convergent 
or asymptotic to the expanded function $f(1/Q^2)$: our approach is more 
general and can be applied whenever the remainder $R_{n}(1/Q^2)$ obeys 
(\ref{V2}) at least for {\it one fixed} value of $n$, i.e., tends to zero 
as $1/Q^{2n}$ or faster in the euclidean domain. This allows, under the 
conditions specified below, an analytic continuation of the truncation error 
from the euclidean to the Minkowski domain. If the starting inequality 
(\ref{V2}) is known for {\it several} values of $n$, one can perform the 
continuation for each of them, term by term. 

Recent applications of the operator product expansion in QCD have focused on
problems dealing with quantities that are essentially related to the Minkowski 
domain, where the properties of the OPE are least known and may be completely 
different from those in the Euclidean domain. Inclusive decays of heavy 
flavours have been discussed. The 't Hooft model \cite{Hoom} of two-dimensional 
QCD in the limit of many colours has been considered, \cite{ChibShif}, 	
with the aim to abstract general features that may survive in four-dimensional 
QCD. We refer the reader to the papers quoted for details. Here let us 
briefly discuss the example of the determination of 
$\alpha_{s}(m^{2}_{\tau})$ from the $\tau$ lepton hadronic 
width, to illustrate physical relevance of the problem.

\section{Determination of $\alpha_{s}(m_{\tau}^{2})$ from the
$\tau$ lepton hadronic width}

The $\tau$ lepton occupies a special position among all leptons, being 
the only lepton heavy enough to decay into hadrons. The $\tau$ decay
provides a unique chance to study hadronic weak interactions at 
moderate energies. There has been extensive interest in using 
measurements of its total hadronic decay width $R_{\tau}$ 
(normalized to the leptonic width),
\begin{equation}
R_{\tau} = \frac{\Gamma[\tau^{-} \rightarrow \nu_{\tau}
+ {\rm hadrons}]}{\Gamma[\tau^{-} \rightarrow 
\nu_{\tau}{\rm e}^{-}\bar{\nu}_{\rm e}]} ,
\label{Rta1}
\end{equation}
to extract the renormalized strong-coupling parameter $\alpha_{s}$. 
This quantity possesses a number of advantages compared with other QCD
observables. It is expected to be calculable in QCD using analyticity and 
the operator product expansion. It is an inclusive quantity which has been
calculated perturbatively to the order $O(\alpha_{s}^{3})$. The $\tau$ mass,
big as it is, is nevertheless below the threshold for charmed hadron
production. 	

Starting from analyticity and the operator product expansion Braaten, 
Narison and Pich \cite{BNP} used the measurements of the $\tau$ decay 
rate to determine the QCD running coupling constant at the scale of the 
$\tau$ mass $m_{\tau}$.	The ratio $R_{\tau}$ is represented in the form
\begin{equation}
12 \pi \int_{0}^{m_{\tau}^2} (1-s/m_{\tau}^2)^2 
(1+2s/m_{\tau}^2) {\rm Im}\,\Pi(s)\frac{{\rm d}s}{m_{\tau}^2} ,
\label{Rta2}
\end{equation} 
where $\Pi(s)$ is a combination of correlators $\Pi^{\mu,\nu}_{i,j,V}$,
$\Pi^{\mu,\nu}_{i,j,A}$ corresponding to the two-point 
correlation functions for the vector $V_{ij}^{\mu}=
\bar{\psi}_{j}\gamma^{\mu}\psi_{i}$ and axial vector 
$A_{ij}^{\mu}=\bar{\psi}_{j}\gamma^{\mu}\gamma_{5}\psi_{i}$ 
colour singlet massless quark currents, with coefficients given by 
the elements $V_{{\rm ud}}$ and $V_{{\rm us}}$ of the 
Kobayashi-Maskawa matrix, the subscripts $i, j=$u,d,s 
denoting light quark flavours. For instance, 
\begin{equation}
\Pi^{\mu,\nu}_{i,j,V}(s) = i\int{\rm d}^4 x\,e^{i q\cdot x}\,\langle 0|
\,T\,\{V^{\mu}_{i,j}(x), V^{\nu}_{i,j}(0)^\dagger\}\,|0\rangle
   = (q^\mu q^\nu - g^{\mu\nu} q^2)\,\Pi_{i,j,V}(s).
\label{correlator} 
\end{equation} 
The integral (\ref{Rta2}) cannot at present be calculated from QCD, 
because the hadronic functions are sensitive to the non-perturbative
effects confining quarks in hadrons. But one can make use of the 
analyticity property of the correlating functions in the complex 
$s$-plane cut along the positive real semiaxis. This allows one to 
express (\ref{Rta2}) as a contour integral along the circle of radius
$m_{\tau}^{2}$:
\begin{equation}
6 \pi {\rm i} \oint_{|s|=m_{\tau}^2} (1-s/m_{\tau}^2)^2 
(1+2s/m_{\tau}^2) \Pi(s)\frac{{\rm d}s}{m_{\tau}^2} 
\label{Rtau}
\end{equation}
(see Fig. 1, showing how analyticity is used to obtain $R_{\tau}$ from 
(\ref{Rtau})). While in (\ref{Rta2}) the integration path $0 \leq s 
\leq m_{\tau}^{2}$ 
runs along the cut, the integration contour in (\ref{Rtau}) keeps 
distance from it (with the exception of one point, $s=m_{\tau}^{2}$, and its 
neighbourhood), thereby giving a justification for representing 
$\Pi(s)$ as the operator-product expansion over local 
operators (provided that the value of $m_{\tau}$ is large enough, 
see a discussion in \cite{BNP}). 

In an analogous way, other weighted integrals (moments) of ${\rm Im}\Pi(s)$,  
\begin{equation}
R_{\tau, \, V/A}^{l,\,m}(s_{0}) = \int_{0}^{s_{0}}{\rm d} s
(1-s/s_{0})^{l}(s/m_{\tau}^{2})^{m} \,\,\frac{{\rm d}R_{\tau, \, 
V/A}^{l,\,m}}{{\rm d}s} , 
\label{hmom}
\end{equation}
have approximately been calculated within the framework of QCD. These
moments can also be expressed as contour integrals analogous to
(\ref{Rtau}). 

It is to be expected that the error brought about by truncating the 
operator-product expansion of $\Pi(s)$ will be larger along rays that 
are closer to the cut $s \geq 0$. There is in particular a special danger 
that the integral (\ref{Rtau}) receives essential contributions from an 
interval around $s=m_{\tau}^{2}$, where the OPE has little chance appropriately 
to represent the function expanded. A fortunate circumstance is that the 
double zero of the kinematic factor $(1-s/m^{2}_{\tau})^2$ in the integrand 
suppresses the contribution from this dangerous segment.  But a quantitative 
analysis of this argument is, to the best of our knowledge, still lacking.
Moreover, as is emphasized in \cite{LeDP}, experimental data are, because of 
the same factor  $(1-s/m^{2}_{\tau})^2$, very poor and statistically limited 
around this point. An explicit estimate of the error is therefore desirable.    

It is our aim to develop a scheme allowing a quantitative analysis of 
this qualitative argument. The method of obtaining the value of $\alpha_{s}$ 
from $R_{\tau}$ was repeatedly criticised in the literature 
\cite{Shif,ANRid} for various reasons, claiming that the 
theoretical uncertainty usually quoted is underestimated. Leaving aside 
uncertainties related to the truncation of the perturbation series, we focus 
on two aspects related to the truncation of the operator product expansion. 
These aspects are:

1. The integral in (\ref{Rtau}) is carried out along the circle $|s| = 
m_{\tau}^2$. When $\Pi(s)$ in the integrand is approximated by the 
truncated OPE series, how does the remainder depend on the direction of 
the ray in the complex $s$ plane? One rightly expects that the error will 
increase with $s$ getting closer to the cut. But the quantitative aspect of 
this statement is unclear. How reliable is the expansion at those points 
$|s|= m_{\tau}^2$ that are far from the euclidean interval, yet not quite 
close to the double zero at $s=m_{\tau}^2$ ? 

2. The terms exponential in $(-s)^{1/2}/\Lambda$, possibly to be added to the 
operator product expansion, are  small along the euclidean ray and invisible 
in the truncated series, but may become big when the ray in the $s$ plane 
approaches the cut in the Minkowski domain. How does the effect of such 
additional terms depend on the deflection from euclidean region?

In the subsequent sections, we discuss these problems by examining the
mathematical model outlined in sec. 2. We find conditions for obtaining 
bounds on the truncation errors in the whole complex high-energy 
domain; it turns out that these bounds are larger (and die off slower) 
in the Minkowski domain than in the euclidean domain.


\section{Results and discussion}

According to the notation introduced in Sec. 2, $f(z)$ represents a 
function holomorphic in a disk centred at the origin of the complex 
$z$ plane cut along the negative real semiaxis. 

We consider two generic assumption schemes: 

{\it Case I} : Let us assume that $f(z)$ has the form  of the 
generalized Stieltjes integral  (see \cite{BOrs})
\begin{equation}
f(z)=\int_{0}^{\infty}\frac{\rho(t)}{1+zt}{\rm d}t 
\label{BOrs1} 
\end{equation}
and that the moments 
\begin{equation}
a_{k}=\int_{0}^{\infty} t^k \rho(t){\rm d}t
\label{BOrs2}
\end{equation}
exist for $k=0,1,2,...,n-1$, with $n$ being a fixed positive integer and 
$\rho(t)$ a real-valued function. Then, assuming a bound on the 
remainder (\ref{rem}) in euclidean region, we obtain a bound along all rays
(see the formulae (\ref{BOrs5}) and (\ref{Vr3})). 

{\it Case II} : The integral representation (\ref{BOrs1}) is not a dispersion 
relation, because the function $\rho(t)$ has to vanish very rapidly for the 
moments $a_{k}$ (\ref{BOrs2}) to exist. Beside this, non-power-like terms added 
to the operator-product expansion may violate the conditions (\ref{BOrs1}), 
(\ref{BOrs2}). We therefore build in subsection 4.2 a framework for more 
general situations, assuming an overall constant bound on $R_{n}(z)$. Then,
using (\ref{V2}) on $R_{n}(z)$ along the euclidean ray, we obtain an
angle-dependent bound along all rays (see the formula (\ref{Vrk3}) or  
(\ref{Vrk4})), and observe how it deteriorates when the ray deflects 
from euclidean to the Minkowski region. The result (\ref{Vrk3}), 
(\ref{Vrk4}) is based on Theorem A of Appendix A. 

In either case, the bounds obtained cannot be improved unless the respective 
class of functions is reduced to a smaller one. We show this by giving 
examples of functions saturating them in subsection 4.1 and in 
the Appendix C respectively.

\subsection{Case I : The remainder in the form of a Stieltjes integral}

Let us first assume that (\ref{BOrs1}) and (\ref{BOrs2}) hold with 
$\rho(t)$ nonnegative for $t \geq 0$. The remainders  
\begin{equation}
R_{k}(z) = (-z)^{k}\int_{0}^{\infty}\frac{\rho(t)}{1+zt}t^{k} 
{\rm d}t ,
\label{BOrs4}
\end{equation}
with $k=1,2,...,n-1$, are bounded, for $z$ approaching zero along the 
positive real semiaxis, by 
\begin{equation}
|R_{k}(z)|=\tilde{a}_{k}(z)|z|^k \leq a_{k}|z|^k  ,
\label{BOrs5}
\end{equation}
where we use the notation 
\begin{equation}
\tilde{a}_{k}(z) =
\int_{0}^{\infty}\frac{\rho(t)}{|1+zt|}t^{k}{\rm d}t .   
\label{BOrs6}
\end{equation}
It is easily seen that the inequality (\ref{BOrs5}) is valid also for $z$ 
approaching zero along any ray lying in the right half of the complex plane 
because $|R_{k}(z)| \leq \tilde{a}_{k}(z)|z|^k$ and $\tilde{a}_{k}(z) 
\leq a_{k}$  for Re$z>0$. This, however, is not the case for Re $z < 
0$, due to the presence of the cut in this halfplane. The following bound 
can be obtained for Re$z<0$. We have, denoting 
$z= r{\rm e}^{{\rm i}\varphi}$, 
\begin{equation}
|1+zt| = (1+2rt \cos \varphi + r^{2}t^{2})^{1/2} .
\label{Vr1}
\end{equation}
Considered as a function of $t$ at $t>0$ and $\varphi$ fixed 
($\pi/2 \leq |\varphi| \leq \pi$), $1/|1+zt|$ has its maximum at 
$t=-\frac{1}{r}\cos \varphi$, where its value is 
$1/|\sin \varphi|$. In this way we obtain (\ref{BOrs5}) and 
\begin{equation}
|R_{k}(z)| \leq a_{k}\, |z|^k /|\sin \varphi|
\label{Vr3} 
\end{equation}
for Re$\,z > 0$ and Re\,$z < 0$ respectively. Comparing
(\ref{Vr3}) with (\ref{BOrs5}), we see how the factor 
$1/|\sin \varphi|$ makes the estimate looser when the ray
gets closer to the cut, i.e., when $\varphi \rightarrow \pm \pi$.  
 
The estimates become worse if the discontinuity along the cut is not 
positive definite. The corresponding bounds can be obtained by 
replacing $\rho(t)$ with $|\rho(t)|$ in the derivation. 

Better bounds on $|R_{k}(z)|$ than (\ref{BOrs5}) and (\ref{Vr3}) 
cannot be obtained unless some special assumptions  about 
$\rho(t)$ are made. To see this, choose $\rho(t)=\rho_{0}(t)$ 
such that $f(z)$ has a pole, $\rho_{0}(t) = c \delta(t-t_{0})$ with $c>0$ 
and $t_{0}>0$, in which case $f(z) = c/(1+zt_{0})$, $a_{k} = c t_{0}^{k}$, 
and $|R_{k}(z)| = c |z|^{k} t_{0}^{k}/|1+zt_{0}|^{k}$. While the bound 
(\ref{BOrs5}) for Re$z > 0$ is saturated for $t_{0} > 0$ tending to zero, 
(\ref{Vr3}) for Re$z < 0$ is saturated for $t_{0}=-\frac{1}{r}\cos \varphi$. 

It is of interest to see how the bounds (\ref{BOrs5}) and (\ref{Vr3}) on 
the truncation error may affect the accuracy of the determination of the 
generic contour integral (\ref{Rtau}). As was discussed above, 
(\ref{BOrs5}) and (\ref{Vr3}) are not immediately applicable in QCD to 
estimate the truncation error, due to the simplifying assumptions of our 
model. But it is interesting to compare them with the conventional 
assumption that the error is constant in all directions of the complex $z$ 
plane, which is a cruder approximation. To see this let us 
examine how the factor $|\sin \varphi|^{-1}$ in (\ref{Vr3}) affects the 
estimate of the integrals of the form (\ref{Rtau}).

As we have neglected the logarithmic dependence of the coefficient
functions, the integral (\ref{Rtau}) can be evaluated trivially 
using Cauchy's residue theorem. Representing the integration variable $s$ in 
the form $s=m^{2}_{\tau} {\rm e}^{{\rm i}\chi}$, $\chi = \pi - \varphi$, and 
replacing the function $\Pi(s)$ by the bound (\ref{BOrs5}), (\ref{Vr3}) on 
$R_{k}(z)$, we obtain the following bound on the integrated remainder:  
\begin{equation}
a_{k} 6 \pi m_{\tau}^{-2k} \left( I_{{\rm Eu}}^{2,1}+
I_{{\rm Msin}}^{2,1} \right) , 
\label{bound}
\end{equation} 
where  
\begin{equation} 
I_{{\rm Eu}}^{l,m} = \int_{\pi/2}^{3\pi/2}{\rm d}\chi |1-{\rm
e}^{{\rm i}\chi}|^{l}|1+2{\rm e}^{{\rm i}\chi}|^{m}  
\label{IEu}
\end{equation}
and 
\begin{equation}
I_{{\rm Msin}}^{l,m} = \left(\int_{0}^{\pi/2}+ 
\int_{3\pi/2}^{2\pi}\right) 
{\rm d}\chi |1-{\rm e}^{{\rm i}\chi}|^{l}|1+2{\rm 
e}^{{\rm i}\chi}|^{m}/|\sin \chi| 
\label{IMsin}
\end{equation}
corresponds to the euclidean and the minkowskian halfplane
respectively.  

The effect by which the discontinuity along the cut tells on the value of
the truncation error can be seen when the sum 
\begin{equation}
I_{2}^{l,m} = I_{{\rm Eu}}^{l,m} + I_{{\rm Msin}}^{l,m}
\label{I2} 
\end{equation}
is compared with   
\begin{equation}
I_{1}^{l,m} = I_{{\rm Eu}}^{l,m} + I_{{\rm Mi}}^{l,m} , 
\label{I1}
\end{equation}
where  
\begin{equation}
I_{{\rm Mi}}^{l,m} = \left(\int_{0}^{\pi/2}+ 
\int_{3\pi/2}^{2\pi}\right)
{\rm d}\chi |1-{\rm e}^{{\rm i}\chi}|^{l}|1+2{\rm e}^{{\rm i}
\chi}|^{m} , 
\label{IMi}
\end{equation}
in which the integrand does not contain the factor $1/|\sin \chi|$.  
The three integrals can be written in the form
\begin{equation}
I_{{\rm Eu}}^{l,m} =  \int_{\pi/2}^{\pi} C^{l,m}(\chi)\,\,{\rm d}\chi , 
\label{Eu}
\end{equation} 
\begin{equation}
I_{{\rm Msin}}^{l,m} =  \int_{0}^{\pi/2} C^{l,m}(\chi)/\sin \chi \,\,
{\rm d}\chi ,
\label{Msin}
\end{equation} 
and
\begin{equation}
I_{{\rm Mi}}^{l,m} = \int_{0}^{\pi/2} C^{l,m}(\chi)\,\,{\rm d}\chi  
\label{Mi}
\end{equation} 
respectively, where 
\begin{equation}
C^{l,m}(\chi)= 2^{l/2+m+1}(1-\cos \chi)^{l/2}(5/4 + \cos \chi)^{m/2}.
\end{equation} 

We see that (\ref{bound}) is composed of two factors, $6 \pi a_{k} 
m^{-2k}_{\tau}$ (which depends on the $k$-th moment $a_{k}$ and on the 
radius $m^{2}_{\tau}$ of the integration circle), and 
$I^{2,1}_{2}=I^{2,1}_{{\rm Eu}}+I^{2,1}_{{\rm Msin}}$ (which contains 
the factor $1/|\sin \chi|$). Certainly, $I_{{\rm Msin}}^{l,m} > 
I_{{\rm Mi}}^{l,m}$, which fact signals the zero of the denominator in 
(\ref{BOrs4}). While $I_{2}^{0,1}$ is a divergent integral, $I_{2}^{1,1}$ 
turns out to be greater than $I_{1}^{1,1}$ by almost 24 per cent. This 
difference is 6.7 per cent for $(l,m) = (2,1)$ and further decreases with 
increasing $l$, but increases with increasing $m$ at fixed 
$l$. Details can be seen in Table 1.

\begin{table}
[htb]
\begin{center}
\begin{tabular}{|c|c|c|c|c|} \hline 
&&&&\\
$l \,\, \backslash \,\, m$&1&2&3&4 \\&&&&\\ \hline
&&&&\\
1&1.237&1.34&1.45&1.56\\ \hline
&&&&\\
2&1.067&1.104&1.148&1.194 \\ \hline 
&&&&\\
3&1.026&1.043&1.064&1.089 \\ \hline
&&&&\\
4&1.012&1.020&1.032&1.046 \\ \hline
\end{tabular}
\caption{The ratio $I^{l,m}_{2}/I^{l,m}_{1}$ for different values of $l$ and 
$m$. The ratio indicates how much the discontinuity along the cut affects the 
role of the truncation error in the integral $I_{2}^{l,m}$ when compared with 
$I^{l,m}_{1}$.}
\end{center}  
\end{table}

\subsection{Case II : The remainder bounded by a constant}

For functions $f(z)$ that do not satisfy the conditions of Case I, the
following theorem may be useful.    

Let ${\cal C}$ be the plane of complex numbers. Let ${\cal C}(d, \alpha)$,  
$0 < \alpha \leq \pi$, be the open segment of angle $\alpha$, 
of the disk of radius $d>0$ centred at the origin.  
In other words, let ${\cal C}(d, \alpha)$ represent $z=r{\rm e}^{{\rm 
i}\phi}, \,\, |\phi|<\alpha, \,\, 0<r<d$ (see Fig. 2).

\begin{figure}[htb]
  \begin{center}
    \includegraphics*[width=8cm]{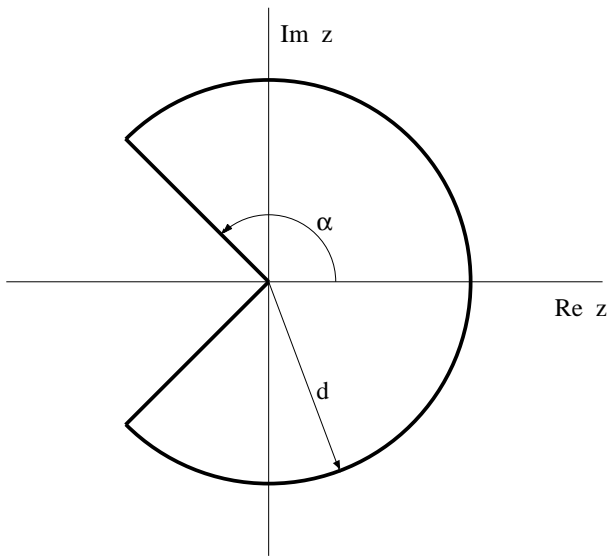}
    \caption{The region ${\cal C}(d, \alpha)$ in the complex $z$ plane. In this
    subsection, the case $\alpha = \pi$ is considered. See the Appendix for the
    general case.}
    \label{fig:t2}
  \end{center}
\end{figure}

{\it Theorem 1.}  Let a function $f(z)$ be holomorphic in 
${\cal C}(d, \pi)$. Let the remainder $R_{n}(z)$, see (\ref{rem}), 
fulfill, for a fixed positive integer $n$, the following two 
inequalities:  
\begin{equation}
|R_{n}(z)| \leq M
\label{Vrk1}
\end{equation}
for complex $z$, $z \in {\cal C}(d, \pi)$,  and 
\begin{equation}
|R_{n}(x)| \leq A x^n 
\label{Vrk2}
\end{equation}
for $0<x<d$. Assume $d = \sqrt[n]{M/A}$. Then, for every 
$\varphi$, $|\varphi| < \pi$, $R_{n}(z)$ satisfies the inequality 
\begin{equation}
|R_{n}(r{\rm e}^{{\rm i}\varphi})| \leq M^{1-\gamma} A^{\gamma} 
\,\, r^{\gamma n}  
\label{Vrk3}
\end{equation}
for all $z \in {\cal C}(d, \pi (1 - \gamma))$, where 	
and $\gamma = 1 - |\varphi|/\pi$. 

This result is a special case of Theorem 2 (and Corollary 1), which is 
proved in the Appendix A. It gives an upper bound on $R_{n}(z)$ along 
every ray passing through the origin, the estimate becoming worse with 
increasing deflection from the positive real semiaxis, i.e., with 
the ray approaching the cut.   

{\it Remark 1.} The equality $d = \sqrt[n]{M/A}$ is a special assumption 
saying that, for $z$ approaching the boundary point on the positive real 
semiaxis, $z=d$, the bounds (\ref{Vrk1}) and (\ref{Vrk2}) coincide. We make 
this assumption for simplicity of the subsequent discussion; 
see Appendix A for the general case. 

{\it Remark 2.} The inequality (\ref{Vrk3}) can also be written in the form
\begin{equation}
|R_{n}(r{\rm e}^{{\rm i}\varphi})| \leq M \, (r/d)^{\gamma n} , 
\label{Vrk4}
\end{equation}
which reveals how the bound deteriorates with increasing distance from the
origin and/or increasing angle (i.e., when energy decreases and,
respectively, when the cut is approached; see a discussion below). 

{\it Remark 3.} Compared with the conditions required in the previous 
subsection, no integral representations for $f(z)$ or the coefficients 
$a_{k}$ are required in the Theorem. On the other hand, by (\ref{Vrk1}) a 
constant bound is imposed on $f(z)$ in ${\cal C}(d, \pi)$, which condition 
was not required in the Case I. In both cases, the resulting bound depends 
on the angle along which infinite energy is approached: In Case I, see  
(\ref{BOrs5}) and (\ref{Vr3}), the coefficient is angle-dependent and the 
exponent $k$ is not, whereas in Case II, see (\ref{Vrk3}) or (\ref{Vrk4}), 
the exponent of $|z|$ is angle-dependent, decreasing from $n$ (along the 
positive real semiaxis) down to 0 (along the cut).  

{\it Remark 4.} Since the resulting estimate (\ref{Vrk3}) in ${\cal C}(d, 
\pi)$ is considerably looser than (\ref{Vrk2}) for $0 < x < d$ (note that 
the exponent is changed from $n$ in (\ref{Vrk2}) to $\gamma n$ in 
(\ref{Vrk3}), $0 \leq \gamma \leq 1$, $\gamma$ tending to zero near the 
cut), it is interesting to look for a function that saturates it. A set 
of functions saturating (\ref{Vrk3}) for different nonnegative integers 
$n$ can be generated by using the function $f(z) = \exp \{{\rm i}a \,\, 
[\, \ln (\sqrt{z} \,\, {\rm e}^{-{\rm i}\pi/2}) \, ]^2\}$ with $a$ real. 
This example (see the Appendix C for details) shows that the bounds are 
optimal, in the sense that they cannot be improved within the class of 
functions considered. There might be physical reasons, on the other 
hand, to restrict oneself to a smaller class of functions, in 
which case an improvement of the bound would be possible. 

The resulting inequality (\ref{Vrk3}) can be used to estimate, under the
assumptions made, the error caused in (\ref{Rtau}) due to approximating 
$\Pi(s)$ in (\ref{Rtau}) by the first $n$ terms of the expansion. We 
observe the following facts:

1. To obtain a bound on the integral (\ref{Rtau}), we assume both 
(\ref{Vrk1}) and (\ref{Vrk2}). The inequality (\ref{Vrk2}) alone (valid only 
in the euclidean region) is not sufficient for obtaining a bound on the 
remainder $R_{n}(z)$ at complex $z$, unless (\ref{Vrk1}) is simultaneously
used. 

2. By Theorem 1, the bounds (\ref{Vrk1}) and (\ref{Vrk2}) are combined to 
create a third one, (\ref{Vrk3}). While (\ref{Vrk1}) is valid in the whole 
complex region ${\cal C}(d, \pi)$ but is loose, (\ref{Vrk2}) holds only 
on the segment $0<x<d$ but is far more restrictive on this interval. 
Theorem 1 combines them to produce (\ref{Vrk3}), which holds in 
${\cal C}(d, \pi)$ {\it and}  is restrictive, although less than 
(\ref{Vrk2}), becoming (\ref{Vrk2}) and (\ref{Vrk1}) on the positive and 
the negative real semiaxis respectively. The resulting bound $M^{1-\gamma} 
A^{\gamma}\,r^{\gamma n}$ on the right hand side of (\ref{Vrk3}) becomes 
(\ref{Vrk2}) and (\ref{Vrk1}) for $\gamma=1$ (euclidean region) and 
$\gamma$ approaching zero (minkowskian region) respectively.

3. For $r=d$, (\ref{Vrk3}) is no improvement of (\ref{Vrk1}). For $r < d$, 
(\ref{Vrk3}) does imply an improvement of (\ref{Vrk1}) thanks to (\ref{Vrk2}), 
by means of the factor $(r/d)^{\gamma n} = r^{\gamma n} (A/M)^{\gamma}$ on 
the right hand side. This factor is smaller than 1 for $r<d$ and 
$|\varphi| < \pi$, but approaches 1 when $|\varphi|$ approaches $\pi$ at 
fixed $r$, or when $r$ approaches $d$ at fixed $\varphi$.    

4. These results induce analogous relations between the corresponding 
integrals of the type (\ref{Rta2}), (\ref{Rtau}) and (\ref{hmom}). Let us 
introduce the notation $\rho = r/d$. Inserting $M \rho^{n(1-|\varphi|/\pi)}$, 
the right-hand side of (\ref{Vrk3}), into (\ref{Rtau}) instead of $\Pi(s)$, 
we find that the integral is bounded, due to (\ref{Vrk1}), by 
\begin{equation}
M I^{l,m}_{1}
\label{RTa1}
\end{equation}
while (\ref{Vrk3}) imposes the bound 
\begin{equation}
M J^{l,m}(n,\rho)
\label{RTa2}
\end{equation}
with 
\begin{equation}
J^{l,m}(n,\rho)=\int_{0}^{\pi}C^{l,m}(\chi)\,\,\rho^{n \chi/\pi}\,\,
{\rm d}\chi  . 
\label{RTa3} 
\end{equation}
on the same integral.

\begin{table}
[htb]
\begin{center}
\begin{tabular}{|c|c|c|c|c|c|c|c|} \hline 
&&&&&&&\\
$n \,\, \backslash \,\, \rho$&0.05&0.2&0.4&0.6&0.8&0.9&1 \\&&&&&&&\\ \hline
&&&&&&&\\
1&0.18&0.38&0.57&0.73&0.87&0.936&1 \\ \hline
&&&&&&&\\
2&0.051&0.17&0.34&0.54&0.76&0.87&1 \\ \hline 
&&&&&&&\\
3&0.020&0.081&0.21&0.40&0.66&0.82&1 \\ \hline
&&&&&&&\\
4&0.009&0.044&0.13&0.30&0.58&0.77&1 \\ \hline
\end{tabular}
\caption{Thanks to (\ref{Vrk2}), (\ref{Vrk1}) improves into (\ref{Vrk3}) and
the corresponding integrated bound $M I^{l,m}_{1}$ into $M J^{l,m}(n,\rho)$. 
The values of $J^{2,1}(n,\rho)/I^{2,1}_{1}$ shown in the table for different
$n$ and $\rho$ illustrate the effect of Theorem 1 on the integrated bounds. 
Small values mean a big improvement, 1 means no improvement.} 
\end{center}  
\end{table} 

As we have seen, (\ref{Vrk2}) can be used to improve the original bound 
(\ref{Vrk1}) into (\ref{Vrk3}). This induces an improvement of the bounds on 
the corresponding integrals, changing $M I^{l,m}_{1}$ into $M J^{l,m}(n, \rho)$. 
The improvement is pronounced for small values of $\rho$ (which means that the 
disk ${\cal C}(d, \pi)$ is large), but becomes weak for $\rho$ approaching 1, 
when the boundary circle of ${\cal C}(d, \pi)$ is approached. The values of the 
ratio $J^{l,m}(n, \rho)/I^{l,m}_{1}$ for $(l,m) = (2,1)$ and some typical $n$ 
and $\rho$ are shown in Table 2. 

It is difficult to make a straightforward comparison of the two bounds, 
(\ref{Vrk3}) on one side and (\ref{BOrs5}), (\ref{Vr3}) on the other, 
because they have been derived under different conditions. Perhaps 
the most striking difference is that, in Case I, $R_{k}(z)$ is
holomorphic in the whole cut $z$ plane, while in Case II it is 
holomorphic only in the cut disk ${\cal C}(d, \pi)$; in this way, 
singularities of any kind are allowed outside ${\cal C}(d, \pi)$, 
arbitrarily near the boundary circle. The two assumption schemes 
therefore assign similar analyticity properties to $f(z)$ if the disk 
${\cal C}(d, \pi)$ is very large, i.e., for small values of $\rho$ 
in (\ref{RTa3}). 	
\begin{table}
[htb]
\begin{center}
\begin{tabular}{|c|c|c|c|c|c|c|c|} \hline 
&&&&&&&\\
$n \,\, \backslash \,\, \rho$&0.05&0.2&0.4&0.6&0.8&0.9&1 \\&&&&&&&\\ \hline
&&&&&&&\\
1&3.7&1.9&1.43&1.21&1.086&1.040&1 \\ \hline
&&&&&&&\\
2&20&4.1&2.1&1.49&1.18&1.081&1 \\ \hline 
&&&&&&&\\
3&155&10&3.3&1.9&1.29&1.13&1 \\ \hline
&&&&&&&\\
4&$10^3$&27&5.2&2.3&1.41&1.17&1 \\ \hline
\end{tabular}
\caption{When extrapolated into the cut disk, (\ref{Vrk2}) deteriorates into
(\ref{Vrk3}) and the corresponding integrated bound $M \rho^{n}I^{l,m}_{1}$ 
into $M J^{l,m}(n,\rho)$. The table shows the values of 
$J^{2,1}(n,\rho)/(\rho^{n}I^{2,1}_{1})$ for different $n$ and $\rho$. Large 
values, which signal a big deterioration (due to a strong angle dependence of 
(\ref{Vrk3})), appear near the origin (where $\rho$ is small) and/or at high 
expansion orders $n$.} 
\end{center}  
\end{table} 

The constant $M$ in (\ref{Vrk1}) affects the value of the integral bound 
(\ref{RTa2}) in combination with the damping factor $\rho^{n \chi/\pi}$ in 
the integrand of (\ref{RTa3}), which becomes small when $n$ increases and/or 
$\rho$ decreases. It is instructive to compare the resulting bound 
$M J^{l,m}(n, \rho)$ with that based on the assumption that (\ref{Vrk2}) 
preserves its form in the whole cut disk. A convenient way to measure 
this effect is to divide $M J^{l,m}(n, \rho)$ by $M \rho^{n} I^{l,m}_{1}$, 
in which the angle-dependent factor $\rho^{n \chi/\pi}$ in the integrand is
suppressed, the exponent $n \chi/\pi$ being replaced by 
its value along the euclidean ray, $\chi = \pi$. By this, (\ref{Vrk2}) is 
extended onto the whole cut disk. The resulting ratio is equal to 
$J^{l,m}(n, \rho)/(\rho^{n} I^{l,m}_{1})$; some numbers are given in Table 3 
to illustrate the effect. While Table 2 shows what improvement of (\ref{Vrk1}) 
is achieved thanks to (\ref{Vrk2}) and the Theorem 1, Table 3 shows how 
strong the standard assumption of an overall validity of (\ref{Vrk2}) is. Out of 
the three bounds, $M\rho^{n}I^{l,m}_{1}$ is the most restrictive one, while our 
result $M J^{l,m}(n, \rho)$ based on the Theorem 1 is looser (as is shown 
in Table 3), and $M \,I_{1}^{l,m}$ based on (\ref{Vrk1}) is the loosest (as is 
seen from Table 2). Thus, 
\begin{equation}
M \, \rho^{n}\, I^{l,m}_{1} \leq M \, J^{l,m}(n, \rho) \leq
M \, I_{1}^{l,m} .
\end{equation}

The efficiency of our bound depends on the value of $\rho$, the 
ratio of $1/d$ (the lowest value of $|s|$ at which the bound (\ref{Vrk1}) is 
supposed to hold) to $m_{\tau}^{2}$ (the radius of the integration contour of 
(\ref{Rtau}) in Fig. 1). Clearly, the problem has sense only for $\rho < 
1$, when the integration contour lies in the analyticity region. If the input 
bounds (\ref{Vrk1}) and (\ref{Vrk2}) are valid down to very low energies 
(i.e., if $1/d$ is small with respect to $m_{\tau}^{2}$), $\rho$ will be small 
and the right-hand side of (\ref{Vrk3}) will become small as well, thanks to 
the factor $\rho^{n \chi/\pi}$. This improving factor is, however, strongly 
angle-dependent in the complex $s$ plane and will eventually, near the cut, 
rise to unity, thereby raising the bound (\ref{Vrk3}) back to the 
starting inequality (\ref{Vrk1}) at the points of the cut.  

These results will be modified when the logarithmic dependence of the 
coefficient functions $a_{k}(q)$ is taken into account. 


\section{Concluding remarks}

It has been our aim to propose a framework allowing a quantitative estimate 
of the truncation error of the operator-product expansion away from euclidean
region. In considering the evaluation of the contour integrals of the type 
(\ref{Rta2}, \ref{Rtau}) or (\ref{hmom}), we have proposed  two different sets 
of model assumptions to estimate the influence of the cut $s>0$ on the 
truncation error along different rays in the complex $s$ plane. In either case, 
the starting relation is the inequality (\ref{V2}) for negative $s$, $s<-b<0$. 
When combined with analyticity, (\ref{V2}) can be extended into the complex $s$ 
plane, but additional assumptions are necessary. We have cosidered two sets 
of them, (\ref{BOrs1}), (\ref{BOrs2}) (see Case I of section 4), and 
(\ref{Vrk1}) (see Case II). The assumptions are rather strong in both 
cases, but are weaker than a straightforward extension of the inequality 
(\ref{V2}) into the $s$ plane. Also the resulting estimates ((\ref{BOrs5}), 
(\ref{Vr3}) and, respectively, (\ref{Vrk3})) are considerably looser 
than such a straightforward extension. Some examples to 
illustrate this are in Table 3. 

This result indicates, in view of the fact that our bounds can be saturated, 
that in applying an operator product expansion one should not mechanically 
extend (\ref{V2}) into the complex $s$ plane without special physical 
justification. 

In either case, the resulting bounds exhibit a pronounced angle dependence in 
the $s$ plane, becoming worse and worse with increasing deflection from the
euclidean region down to the cut. As a consequence, the integrated bounds on 
(\ref{Rta2}, \ref{Rtau}) or (\ref{hmom}) are looser than those based on the a 
priori assumption of angle independence.  
This result can be understood as a warning that conventional truncation error 
estimates used in phenomenology are perhaps too optimistic. 

As our results are valid for certain classes of functions, 	
a reduction of the class considered to a smaller one could yield a more
restrictive upper bound. It is a challenge for physics to look for physically 
motivated class reductions. 	 

Our results do not represent the complete solution to the problem. As mentioned
in the Introduction, it was our aim to propose a model scheme that ({\it i}) 
would have precise mathematical meaning, ({\it ii}) would be free from the 
{\it a priori} assumption that the discontinuity along the cut has no 
influence on the truncation error along a general ray and, simultaneously,  
({\it iii}) would keep the model possibly close to real situations. Whereas 
our model scheme satisfactorily meets the conditions ({\it i}) and ({\it 
ii}), it does not do full justice to the requirement ({\it iii}). We have 
made a step out of the schemes based on the a priori assumption that the 
truncation error is the same along all rays $|s| \rightarrow \infty$ passing 
through the origin. Our approach has been based on a plausible, but still 
crude picture of the operator product expansion.  Further refinement 
is necessary. Next step should include allowance for the logarithmic 
$q$--dependence of the coefficients $a_{k}(q)$, and also a discussion of 
current QCD models \cite{Hoom}, \cite{ChibShif}, \cite{deR}. An improvement of 
the integrated error estimate may be reached by introducing a subtracted 
dispersion relation and reversing the order of the $s$-integration and the 
$t$-integration (inclusive processes). Work along these lines is in progress.

{\bf Acknowledgements:} We thank I. Caprini, E. de Rafael, A. Kataev, P. 
Kol\'{a}\v{r}, V. Smirnov and J. Stern for stimulating discussions, and M.
Flato for drawing our attention to Ref. \cite{Fer}. One of us (J.F.) is 
indebted to A. de R\'{u}jula and the CERN Theory Division for hospitality. 
We acknowledge the support of the grants Nos. GAAV-A1010711, MSMT-VS96086, 
and GACR-202/96/1616. 

\appendix 
\section{Appendix: Theorem A and its proof} 

 Let ${\cal C}$ be the plane of complex numbers and ${\cal C}(d, \alpha) 
 = \{ z \in C : z = r{\rm e}^{{\rm i}\varphi}, |\varphi|< \alpha, 0<r<d\}$ 
 where $d>0$ and $0<\alpha \leq \pi$ are given (see Fig 2). 

{\bf Theorem 2}. Let $f(z)$ be holomorphic in ${\cal C}(d , \alpha)$
and fulfilling 
\begin{equation}
|f(z)| \leq M
\label{VrA1}
\end{equation}
for $z \in {\cal C}(d, \alpha)$ and 
\begin{equation}
|f(x)| \leq N x^{n} 
\label{VrA2}
\end{equation}
for $0 < x < d$, where $M$ and $N$ are positive constants. Denote 
$d_{0}= \sqrt[n]{M/N}$. Then  
\begin{equation}
|f(z)| \leq |z|^{n/2} \sqrt{NM} 
\label{VrA3}
\end{equation} 
for $z \in {\cal C}(d, \alpha/2)$, $|z| \leq \min (d_{0}, d)$ and 
\begin{equation}
|f(z)| \leq M  \,\,\,\,\,\,\,\,\,\,\,\, for \,\, z \in {\cal C}(d, \alpha/2),
\,\,\,\,\,\, \,\,\,\,
\min (d_{0}, d) \leq |z| \leq  d . \\
\label{VrA4}
\end{equation} 
{\bf Remark 1.} Repeating the statement of the Theorem we conclude
that under the conditions of the Theorem we have: for every
nonnegative integer $s$ 
\begin{equation}
|f(z)| \leq {\hat N}_{s} |z|^{n/2^s} 
\label{VrA31}
\end{equation} 
for $z \in {\cal C}(d, \alpha(1 - 2^{-s}))$, $|z| \leq 
\min (d_{0}, d)$ and 
\begin{equation}
|f(z)| \leq M  \,\,\,\,\,\,\,\,\,\,\,\, for \,\, z \in 
{\cal C}(d, \alpha(1 - 2^{-s})),
\,\,\,\,\,\, \,\,\,\,
\min (d_{0}, d) \leq |z| \leq  d ,  \\
\label{VrA41}
\end{equation} 
where 
\begin{equation}
{\hat N}_{s} = M(N/M)^{1/2^s} 
\label{VrA51}
\end{equation}
is valid. This means that the remainder $R_{n}(z)$ (\ref{rem})
is bounded in the whole region ${\cal C}(d, \alpha)$ but the 
estimates are bad near the boundary.  

{\bf Remark 2.} By a reasoning similar to that used in the Proof of
the Theorem the following statement can be proved. If a function
$f(z)$ fulfils the assumptions of the Theorem, then 

\begin{equation}
|f(z)| \leq {\tilde N}_{s}|z|^{n(1-2^{-s}})  
\label{VrA32}
\end{equation} 
for $z \in {\cal C}(d , \alpha/2^{s})$, $|z| \leq \min (d_{0}, d)$ and 
\begin{equation}
|f(z)| \leq M  \,\,\,\,\,\,\,\,\,\,\,\, for \,\, z \in {\cal C}(d_{0}, 
\alpha/2^{s}), \,\,\,\,\,\, \,\,\,\,
\min (d_{0}, d) \leq |z| \leq  d . \\
\label{VrA42}
\end{equation} 
is valid where 
\begin{equation}
{\tilde N}_{s} = M(N/M)^{1-1/2^s} .
\label{VrA52}
\end{equation}
This means that, in a small angle, the estimate can be improved.

Combining these two remarks we have

{\bf Corollary.} If a function $f(z)$ fulfils the assumptions of the
Theorem, then for every $\xi$, $0 < \xi < 1$, the following inequalities
\begin{eqnarray}
|f(z)| \leq N_{\xi} |z|^{n(1-\xi)} \,\,\,\,\,\, 
for \,\,\,\,\,\,\,\, z \in 
{\cal C}(d, \alpha \xi), \,\,\,\,\,\,\, |z| \leq \min (d_{0}, d)  \nonumber \\ 
|f(z)| \leq M  \,\,\,\,\,\,\,\,\,\,\,\, 
for \,\,\,\,\,\,\,\, z \in 
{\cal C}(d, \alpha \xi),
\,\,\,\,\,\, \,\,\,\,
\min (d_{0}, d) \leq |z| \leq  d 
\label{VrA6}
\end{eqnarray} 
are valid where 
\begin{equation}
N_{\xi} = M(N/M)^{1-\xi} .
\end{equation}

{\bf Proof of the Theorem}. Denote
\begin{equation}
f_{1}(z) = \ln |f(z)| - \ln M .
\label{VrA7}
\end{equation}
The symbol $f_{1}(z)$ is understood as the harmonic function 
$f_{1}(x,y)$, where $z=x+{\rm i}y$. The symbol $f_{1}(x)$ is 
understood as $f_{1}(x,0)$. 
$f_{1}$ is a harmonic function in the region ${\cal C}(d, \alpha) -
{\cal N}$ where ${\cal N} = \{z: f(z) = 0\}$ is a countable set without
condensation points in ${\cal C}(d, \alpha)$. The function $f_{1}(.)$
fulfils
\begin{equation}
f_{1}(x) \leq \min (n \ln x + \ln N - \ln M, 0) 
\label{VrA8}
\end{equation}
for real $x$ in $(0, d) - {\cal N}$. Further, 
\begin{equation}
f_{1}(z) \leq 0 
\label{VrA9}
\end{equation} 
in ${\cal C}(d, \alpha)$ with the exception of countably many 
isolated points. 
Define 
\begin{equation}
G(x) = \min(n \ln x + \ln N - \ln M , 0).
\label{VrA10}
\end{equation}
The estimate $G(x)$ of $f_{1}(x)$ is nonpositive for $0 < x \leq d$.

Now we introduce the notation 
${\cal D} =  \{ z \in C : z = r{\rm e}^{{\rm i}\varphi}, 
0 < \varphi < \alpha, 0<r<d\}$, and define the 
function $g_{1}(z)$ which is \\
(a) {\it harmonic, nonpositive and maximal in the region ${\cal D}(d,
\alpha)$ and continuously extensible on $\{x : 0 < x \leq d\}$, a part
of the boundary of  ${\cal D}(d, \alpha)$, with exception of countably
many isolated points}, and  \\
(b) {\it fulfilling}
\begin{equation} 
g_{1}(x) \leq G(x) \,\,\,\,\,\,\,\, {\it for} \,\,\,\, 0 < x \leq d.
\nonumber 
\end{equation}
Such function really exists (see Appendix B). It follows that the 
function fulfils 
\begin{eqnarray}
g_{1}(x) = G(x) \,\,\,\,\,\, for \,\,\,\,\,\, 0 < x \leq d , \\
\nonumber
g_{1}(x \exp({\rm i}\alpha)) = 0 \,\,\,\, for \,\,\,\, 0 < x \leq d , \\
\nonumber
g_{1}(d \exp({\rm i}\varphi)) = 0 \,\,\,\,\, for \,\,\,\, 0 \leq \varphi  
\leq \alpha . 
\nonumber 
\end{eqnarray}
In a similar way, we define a symmetric harmonic function $g_{2}(z)$ 
fulfilling (a) and the inequality
\begin{equation}
g_{2}(x{\rm e}^{{\rm i}\alpha}) \leq G(x) \,\,\,\, for \,\,\,\, 0 < x 
\leq d 
\nonumber 
\end{equation} 
with exception of countably many isolated points. This 
function fulfils 
\begin{eqnarray}
g_{2}(x{\rm e}^{{\rm i}\alpha}) = G(x) \,\,\,\,\,\, for \,\,\, 0 < x \leq d \\
\nonumber
g_{2}(x) = 0 \,\,\,\,\,\, for \,\,\,\,\,\,\,\, 0 < x \leq d \\
\nonumber 
g_{2}(d {\rm e}^{{\rm i}\varphi}) = 0 \,\,\,\,\,\, for \,\,\,\, 
0 \leq \varphi \leq \alpha .
\nonumber
\end{eqnarray}

The sum $g(z) = g_{1}(z) + g_{2}(z)$ certainly fulfils the inequality 
(the $g_{i}$ are maximal)
\begin{equation}
g(z) \leq  G(|z|) 
\nonumber
\end{equation}
so that the symmetry of the functions $g_{i}$ yields 
\begin{equation}
g_{1}(x{\rm e}^{{\rm i}\alpha/2})=
g_{2}(x{\rm e}^{{\rm i}\alpha/2}) \leq G(x)/2 .
\nonumber
\end{equation}
If we compare the functions $f_{1}$ and $g_{1}$ we obtain (see 
Appendix B) 
\begin{equation}
f_{1}(z) \leq g_{1}(z) \,\,\,\,\,\, in \,\,\,\,\,\, {\cal D}(d, 
\alpha) - {\cal N} .
\label{VrA11}
\end{equation}
We have 
\begin{equation}
g_{1}(x{\rm e}^{{\rm i}\alpha/2}) \leq  G(x)/2 .
\nonumber
\end{equation} 
Since $G(x)$ is nonpositive we have 
\begin{equation}
g_{1}(x) \leq G(x)/2 .
\nonumber
\end{equation} 
These conditions together with (\ref{VrA11}) and the maximality yields 
\begin{equation}
f_{1}(z) \leq G(|z|)/2  \,\,\,\,\,\, in \,\,\,\,\,\, {\cal D}(d, \alpha/2) 
\nonumber
\end{equation} 
under the condition that we define $f_{1}(z) = -\infty$ for $z \in 
{\cal N}$. Using the definitions of the functions $f_{1}$ we obtain 
the statement of the Theorem. 

\section{Appendix: Proof of two statements}
(i) {\bf Proof of maximality}. 
The region ${\cal D}(d, \alpha)$ can be conformally mapped on 
the unit disk. Denote $\Gamma$ and $\gamma(.)$ the image of 
the interval $(0, d]$ and the image of the function $G(x)$, 
respectively. Due to the theorem \cite{Hof} (Chap. 3, page 33-34, 
item (v))\footnote{A harmonic function $f$ in the open unit disc is 
the Poisson integral of a finite positive Baire measure if and only if 
$f$ is non-negative.} 
the maximal harmonic function fulfilling the 
condition (a) is given by the Poisson formula
\begin{equation}
\frac{1}{2 \pi} \int_{\Gamma}\gamma(\psi)\frac{1-|z|^2}{1-2|z|\cos(\arg z 
- \psi) + |z|^2}{\rm d}\psi .
\nonumber
\end{equation}

(ii) {\bf Proof of (\ref{VrA11})}. 
Let $\{d_{n}\}$, $\{\alpha_{n}\}$ and $\{\rho_{n}\}$ be sequences 
fulfilling the inequalities $d_{n}<d$, $\alpha_{n} < \alpha$ and 
$\rho_{n}>0$ and converging to $d$, $\alpha$ and 0, respectively. 
Denote ${\cal D}_{n, m} = {\cal D}(\rho_{m}, d_{n}, \alpha_{n}) 
= {\cal D}(d_{n}, \alpha_{n}) - {\cal D}(\rho_{m},\alpha_{n})$. 
Assume that the sequences are chosen so that $\partial{\cal D}_{n,m}
\cap{\cal N}\subset(\rho_{m},d_{n}]$, where $\partial{\cal D}_{n,m}$ 
denotes the boundary of ${\cal D}_{n,m}$. This means that 
${\cal D}_{n,m}$ contains only a finite number of points
from ${\cal N}$. ${\cal D}_{n,m}$ can be conformally mapped on the
unit disc (denoted by ${\cal U}$). Further denote $f^{n,m}$ and
$h^{n,m}$ the image of $f_{1}$ and $f/M$ respectively. Then certainly
$f^{n,m} = \ln |h^{n,m}|$. The function $h^{n,m}$ has only a finite
number of 
zero points in ${\cal U}$. Let $\{z_{1}, z_{2}, ..., z_{s}\}$ be the
zero points with multiplicities $\{n_{1}, n_{2}, ..., n_{s}\}$. Since 
$h_{n,m}/\Pi(z-z_{i})^{n_{i}}$ is a nonzero holomorphic function in 
${\cal U}$ the function 
\begin{equation}
F^{n,m}(z) = f^{n,m}(z) - \sum_{i=1}^{s}
n_{i}\ln|\frac{z-z_{i}}{1-z\bar{z}_{i}}|
\end{equation}
is a harmonic function in ${\cal U}$. 
Since the functions $\ln |(z-z_{i})/(1-z \bar{z}_{i})|$ vanish on 
$\partial{\cal U}$ and $f^{n,m}(z)$ are nonpositive, the functions
$F^{n,m}(z)$ are nonpositive, too. 
Due to the theorem from \cite{Hof} there exists a nonpositive   
measure $\mu$ on $\partial{\cal D}$ such that 
\begin{equation}
F^{n,m}(z) = f^{n,m}(z) - \sum_{i=1}^{s}
n_{i}\ln|\frac{z-z_{i}}{1-z\bar{z}_{i}}| =
\frac{1}{2\pi}\int_{-\pi}^{\pi}\frac{1-|z|^2}{1-2|z|\cos(\arg z -
\psi) + |z|^2}{\rm d}\mu(\psi).
\end{equation}
Let $\Gamma^{n,m}$ and $\gamma^{n,m}$ be the image of the set
$(\rho_{m}, d_{n})$ and of the function $G(.)$, respectively. Using
again the fact that the functions $\ln |(z-z_{i})/(1-z \bar{z}_{i})|$ are 
zero on $\partial{\cal U}$ we have $F^{n,m}(z) = f^{n,m}(z) \leq
\gamma^{n,m}(z)$ on $\Gamma^{n,m}$ and 
\begin{equation}
F^{n,m}(z) \leq  \frac{1}{2 \pi} \int_{\Gamma^{n,m}}\gamma^{n,m}(\psi)
\frac{1-|z|^2}{1-2|z|\cos(\arg z - \psi) + |z|^2}{\rm d}\psi .
\end{equation}
Since the functions $\ln|(z-z_{i})/(1-z\bar{z}_{i})|$,
$k=1,2,...,s$,  
are nonpositive in ${\cal U}$ (with exception of the point 
$z_{i}$), we obtain 
\begin{equation}
f^{n,m}(w) \leq q^{n,m}(w) 
\end{equation}
in ${\cal U}$, where the $q^{n,m}(w)$ are  defined by the last integral. 
Denote $\beta^{n,m}$ the preimage of the function $q^{n,m}(w)$.
Certainly we have 
\begin{equation}
f_{1}(z) \leq \beta^{n,m}(z)  \,\,\,\,\,\,\,\, for \,\,\,\,\, z \in 
{\cal D}_{n,m} .
\end{equation}
Due to (i) the functions $\beta^{n,m}(.)$ fulfil the condition (a) 
of the Appendix A in the region ${\cal D}_{n,m}$ and the condition 
$\beta^{n,m}(x) \leq G(x)$ for 
$x \in (\rho_{m}, d_{n}]$. Further, the functions $\beta^{n,m}(x)$ 
fulfil 
\begin{eqnarray}
\beta^{n,m}(x) = G(x) \,\,\,\,for \,\,\,\, \rho_{m} < x \leq d_{n} , \\
\nonumber
\beta^{n,m}(x \exp[{\rm i}\alpha_{n}]) = 0 \,\,\,\,for 
\,\,\,\, \rho_{m} < x \leq d_{n} , \\
\nonumber
\beta^{n,m}(d_{n} \exp[{\rm i}\varphi]) = 0 \,\,\,\,for 
\,\,\,\, 0 \leq \varphi \leq \alpha_{n} , \\
\nonumber
\beta^{n,m}(\rho_{m} \exp[{\rm i}\varphi])  = 0 \,\,\,\,for 
\,\,\,\, 0 \leq \varphi \leq \alpha_{n} . 
\end{eqnarray}
Since the functions $\beta^{n,m}(.)$ are maximal they form a
nonincreasing sequence for $m \rightarrow \infty$ and fixed $n$ in 
${\cal D}(d_{n}, \alpha_{n})$, and are bounded
from below by the maximal function $\beta^{n}$ in ${\cal D}(d_{n},
\alpha_{n})$, which fulfils (a) in ${\cal D}(d_{n}, \alpha_{n})$ and 
$\beta^{n}(x) \leq G(x)$ for $x \in (0, d_{n}]$. Denote $\nu^{n}(z)=
\inf_{m} \beta^{n,m}(z)$  
with fixed $n$; $\nu^{n}(.)$
is a harmonic function in ${\cal D}(d_{n}, \alpha_{n})$, and since
$\beta^{n}(x)$ is maximal we have $\nu^{n}(z) = \beta^{n}(z)$. 
Certainly we have
\begin{equation}
f_{1}(z) \leq \beta^{n}(z)
\end{equation}
for all $n$. This procedure can be repeated for $n \rightarrow
\infty$ and the inequality (\ref{VrA11}) is proved. 

\section{Appendix: A function saturating the bound 
(\ref{VrA3})}
 
The following example shows that the exponent in (\ref{VrA3}) 
cannot be improved.

{\bf Example.} Let $f(.)$ be the function 
\begin{equation}
f(z) = \exp \{{\rm i}a \,\, [\, \ln (\sqrt{z} \,\, {\rm e}^{-{\rm i}\pi/2})
\, ]^2\}
\label{log}
\end{equation} 
where $a$ is a positive constant. The function is holomorphic in 
$C(\infty, \pi)$ and is bounded in $C(1, \pi)$. The function $f(z)$ can be 
rewritten 
\begin{equation}
f(z) = |\sqrt{z}\,\,|\,^{2 a (\pi/2 - \arg \sqrt{z})} \, \exp \{{\rm i}a\,
[(\ln \sqrt{|z|}\,\,)^{2} - (\pi/2 - \arg \sqrt{z}\,)^{2}]\} 
\nonumber
\end{equation}
such that 
\begin{equation}
|f(z)| = |z|^{a(\pi - \arg z)/2} .
\nonumber
\end{equation}
This yields that $f(x)=x^{a\pi/2}$ along the positive real axis and 
converges to zero as $|z|^{a(\pi - \alpha)/2}$ 
along the ray $R_{\alpha}=\{z: \arg z = \alpha, {\rm Re}z>0\}$.  
Compare this with Corollary.

\end{document}